# Large-scale optothermal assembly of colloids mediated by a gold microplate

Vandana Sharma[1], Diptabrata Paul[1], Shailendra K Chaubey[1], Sunny Tiwari[1] and G.V. Pavan Kumar[1,2]

[1] Department of Physics, Indian Institute of Science Education and Research, Pune-411008, India
[2] Center for Energy Science, Indian Institute of Science Education and Research, Pune-411008, India

E-mail: pavan@iiserpune.ac.in

**Abstract**

Light-activated colloidal assembly and swarming can act as model systems to explore non-equilibrium state of matter. In this context, creating new experimental platforms to facilitate and control two-dimensional assembly of colloidal crystals are of contemporary interest. In this paper, we present an experimental study of assembly of colloidal silica microparticles in the vicinity of a single-crystalline gold microplate evanescently excited by a 532 nm laser beam. The gold microplate acts as a source of heat and establishes a thermal gradient in the system. The created optothermal potential assembles colloids to form a two-dimensional poly-crystal, and we quantify the coordination number and hexagonal packing order of the assembly in such a driven system. Our experimental investigation shows that for a given particle size, the variation in assembly can be tuned as a function of excitation-polarization and surface to volume ratio of the gold microplates. Furthermore, we observe that the assembly is dependent on size of the particle and its material composition. Specifically, silica colloids assemble but polystyrene colloids do not, indicating an intricate behaviour of the forces under play. Our work highlights a promising direction in utilizing metallic microstructures that can be harnessed for optothermal colloidal crystal assembly and swarming studies. Our experimental system can be utilized to explore optically driven matter and photophoretic interactions in soft-matter including biological systems such as cells and micro organisms.

Keywords: thermophoresis, optothermal, colloidal assembly, Voronoi decomposition, gold microplate

## 1. Introduction

Optical manipulation of colloids is an active research area with implilcations in light matter interaction studies including optical trapping (1), optical binding (2-4), sub-topics of soft-matter physics such as light-driven assembly (5), swarming (6) and locomotion of biological matter in a fluidic environment (7, 8). Conventionally, optical tweezers have been utilized to trap colloidal particles and biological cells with sizes ranging from nanometre to micrometre scale (9, 10). Optical tweezers can trap objects by using focused laser beams to create deep potential well through an interplay between scattering and gradient forces (2, 3). These methods are generally limited by the number of particles that can be trapped simultaneously. The behaviour of a collective colloidal matter varies from a single particle and gives insight into various basic physical phenomena. Study of colloidal particles on large scale provides the opportunity to study polycrystals (11), glass transition (12-14), jamming (15), packing structures (16, 17) and two-dimensional melting (18). To create large scale assemblies, we have to look beyond the conventional trapping methods and employ additional forces. To this end, various approaches have been utilized such as using plasmonic thin films (19, 20), modulated light fields (21), Lorentz force between plasmons (22), light scattering of trapping laser by nanoparticles (6,

23) and many more. One of the methods to assemble colloids is to utilize their temperature gradient dependent motility. In this context, thermophoresis has been used in variety of studies to trap colloidal objects (8, 24). Thermophoresis is a particle-solvent interfacial phenomenon where particles of different materials can respond differently to an applied temperature gradient (25-28). Recently, thermoplasmonically driven studies have been conducted to establish temperature gradient in the systems (24, 29, 30). This is achieved by excitation of a plasmonic structure, which can absorb part of the light and generate heat due to photon-phonon conversion (31, 32). It can give rise to several thermodynamic processes in the fluid such as thermophoresis, convection (33), bubble formation (34-36) and Marangoni flows (37, 38). Generally, the generation of heat is considered counterintuitive to the process of trapping but it can be harnessed for our advantage. In this context, thermophoresis has been utilized to manipulate multiple particles down to the level of a single particle (39-41).

Herein, we report large scale assembly of colloidal particles by utilizing opto-thermal effects of a single gold microplate. These microplates are chemically as well as lithographically prepared and offer the advantage of generating more heat compared to other structures of the same volume, as we can excite the volume of a two-dimensional flat structure more efficiently (42). We utilize an evanescent configuration to excite a single gold microplate. We show reverse thermophoresis of silica particles which migrate from a colder to a hotter region and compare it with the response of polystyrene particles (PS). A single gold structure can assemble silica particles in a region spatially much larger than its own size, up to 10 times its own dimension. We quantify the temperature distribution through numerical simulations. The assembly is polycrystalline in nature and is quantified through coordination number and hexagonal order parameter. We further investigate the effect of changing the surface to volume ratio (S/V) and particle size on the assembly. The scale of assembly varies as we switch our polarization from in plane (p) to perpendicular (s) to the plane of incidence. We emphasize that the assembly process is thermal gradient driven and can be potentially harnessed for composition-specific assembly and sorting of colloids in optical fields.

## 2. Methods

### 2.1 Sample preparation

Gold microplates were prepared using a method as described in (43). 6 mL of ethylene glycol was heated at $150^0$ C. The temperature was constantly monitored using a thermometer. Once the temperature equilibrated, 1 mL of 0.2 M Chloroauric acid (HAuCl4) was added. After 5 minutes, 3 mL ethylene glycol solution of Polyvinylpyrrolidone (PVP) was injected into the solution. The molar ratio of PVP/Au was kept at 30. The solution was continuously and vigorously stirred for 30 minutes. Afters few minutes, the appearance of shiny, metallic gold in the solution indicated the formation of gold microplates. The solution was brought to room temperature and washed with ethanol and acetone several times. Finally, the solution was rinsed with milli-Q water and dispersed in ethanol for storage. The typical length of a gold microplate was approximately 30 μm shown in the SEM image of figure 1 (b) and the thickness between 100-200 nm (see figure S1 of Supplementary information).

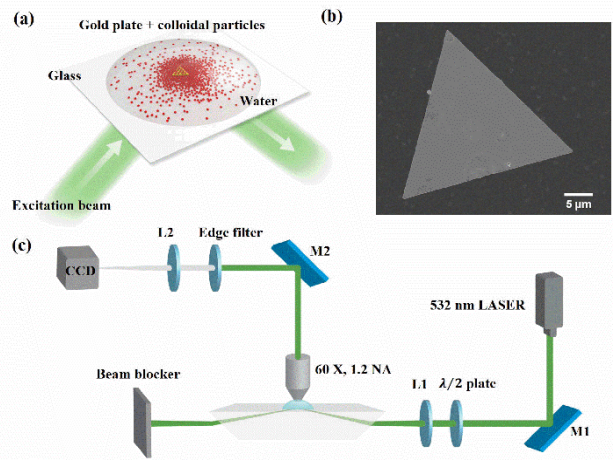

**Figure 1.** **(a)** Schematic illustration of the assembly of colloidal particles on the surface and around the gold microplate. The microplate is shown as a yellow triangular plate and the colloidal particles are indicated by red spheres, which were initially dispersed throughout the solution. The arrows indicate the direction of excitation. **(b)** Field emission scanning electron microscopy (FESEM) image of a representative gold microplate used in the experiments. **(c)** Schematic of the experimental setup: 532 nm laser excitation is incident on the gold microplates. The laser is weakly focused through a 100 mm lens (L1) and undergoes total internal reflection through the dove prism. The video is recorded by a CCD after rejecting the direct excitation using a 532 nm edge filter.

Gold microplates with different S/V ratio were prepared using photolithography. Positive photoresist was spin-coated on glass coverslips. The coverslips were heated at $110^0$C for 1 min. The pattern of the structures was transferred to the substrate by photomask using 405 nm laser followed by immersion in developer solution and cleansing. A 3 nm chromium was coated as an adhesive layer followed by gold coating of desired thickness by thermal vapor deposition.

Colloids ranging from 0.97 μm - 3 μm were used in the experiments. Silica beads (500 nm, 1 μm, 2 μm, 3 μm) were purchased from microsphere-nanosphere. PS particles (0.97 µm) were purchased from Sigma-Aldrich. All the colloidal particles were diluted and dispersed in milli-Q water.

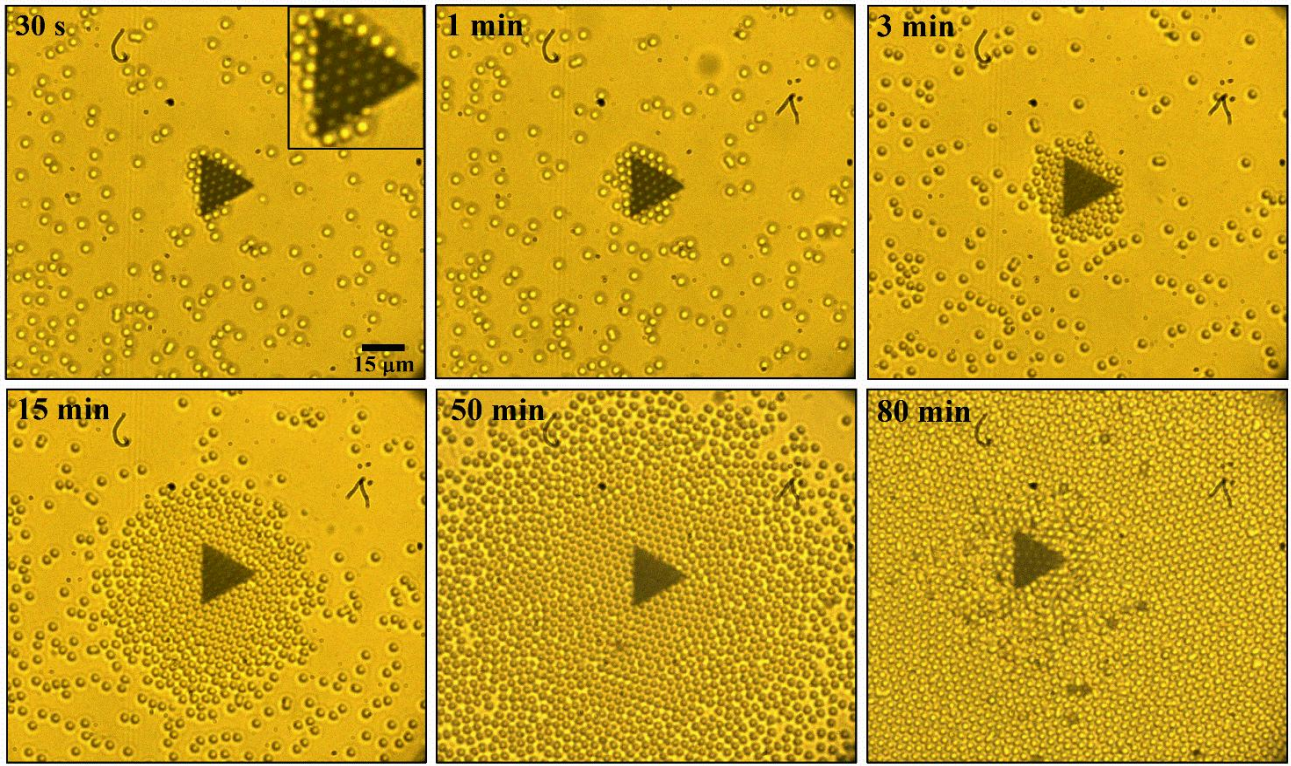

**Figure 2.** Time series images of the gradual accumulation of the 3 µm silica particles around the gold microplate. The figure shows the snapshots of the assembly at different times ranging from 30 seconds to 80 minutes. The inset at 30s shows a zoomed-in image of the gold microplate. The accumulation of the particles on top of the microplate can be clearly seen. The accumulation increases gradually and starts to saturate around 80 minutes.

### 2.2 Experimental details

The gold microplates dispersed in ethanol were dropcasted on a clean coverslip and left to dry. A small volume (30-50 µL) of fluid containing colloids was dropcasted on the coverslip. As shown in figure 1, a 532 nm continuous-wave laser (300 mW) was used to excite the gold plate. A half wave plate was used to switch the polarization form s to p. The laser was loosely focused using a 100 mm lens and was totally internally reflected at the water-glass interface using a dove prism. The size of the evanescent spot is approximately 200 µm. The gold microplate was placed in the excitation region. A white light source was coupled through the bottom of the sample for better visibility. A 60x, 1.20 NA water immersion objective lens was used to capture the scattered light. The scattered excitation laser was rejected using a 532 nm edge filter and the data was captured with a CCD camera.

### 2.3 Numerical simulation details

Temperature distribution and temperature gradient distributions were obtained by solving electromagnetic wave and heat transfer in solids and fluids modules in 3D. Velocity profile of the fluid was obtained by solving for laminar flow in 2D. To get an estimate of the temperature and velocity profiles that could be generated in a micron sized gold triangle, we performed finite element method (FEM) based numerical simulations. Since simulating the exact size of the structure used in the experiments is computationally expensive, we considered a box of dimensions 10 µm x 7 µm x 7 µm. An equilateral gold triangle with side of 2 µm, thickness 50 nm was simulated. All the domains were meshed through user defined values. The structure was illuminated through total internal reflection at 532 nm laser excitation.

## 3. Results and Discussion

### 3.1 Optothermal assembly of silica colloids

We first demonstrate the long-range assembly of 3 µm $SiO_2$ (silica) particles around a triangular gold microplate with each side being ~ 24 µm. The concentration of the particles was 450 particles/ µL. This assembly is facilitated by an opto-thermal potential well created by an evanescent excitation of the gold microplate. As the laser excitation was turned on, we observed a change in the particle dynamics from Brownian motion to a directional movement towards the gold microplate as shown in figure 2.

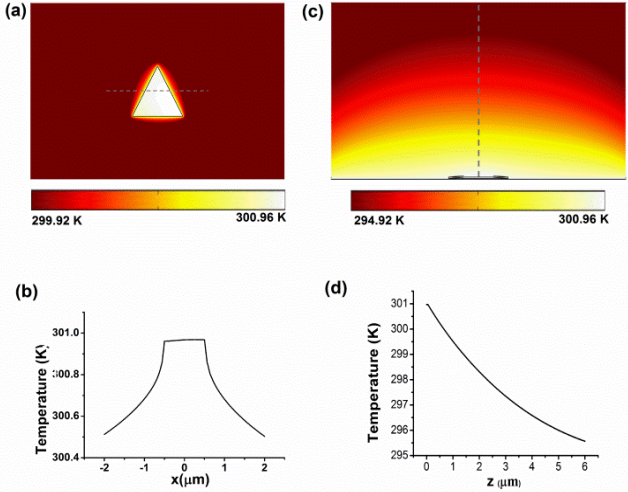

**Figure 3: Finite element method simulations (a)** Temperature profile distribution in x-y plane on the upper surface of the gold plate. **(b)** temperature distribution along the dotted line in (a). **(c)** temperature profile distribution in the x-z plane. **(d)** temperature distribution along the dotted line in (b).

Within the first 30 seconds, the particles get accumulated on the surface of the gold microplate with uniform spacing as shown in the inset of figure 2. It is to be noted that although the particles accumulate on the gold microplate frequently, they do not always assemble in a similar way. As time progresses, more particles started to accumulate around the microplate. The formation of multilayers of the particles can also be observed in the snapshot taken at 80 min within a region close to the microplate, making distinction between the individual particles difficult. A large scale assembly extending up to 292 μm was formed at 80 min (see figure S2 of supplementary information). Thus, a single gold microplate under evanescent excitation is able to accumulate particles up to a distance which is approximately 10 times its own dimension. It is to be noted that since the electric field from the edge of the gold plate decays exponentially, extending up to few hundreds of nm, compared to the particles' assembly size of hundred of microns, the assembly of the particles cannot be facilitated by optical forces alone and is a result of the heat generated due to the structure. We attribute this long-range assembly mainly to thermophoretic migration of silica particles.

In order to explain this assembly process, we can decompose the forces acting on the silica particles, as (a) thermophoretic force and (b) convective force. Thermophoretic force acting on a particle can be either positive or negative depending on the particle-solvent interaction. Convection current on the other hand has a toroidal nature inside the water droplet, and can bring the particles close to the microplate and then transport them away from the plate (33). An interplay between these two forces is crucial in either assembling the particles close to the microplate or diffusing them back into the solution. In the absence of the gold microplate, the only force acting on the particles is the evanescent field of the laser excitation. We did not observe any assembly of the particles in this case, implying that the optical force due to evanescent field of the laser alone does not have any role in the assembly of the particles. When the gold plate is introduced in the same electromagnetic field, there is a redistribution of electronic charges and an induced field is set up inside the structure. The electronic current generated by oscillating electrons dissipates energy via the Joule effect and the heat power density $q(r)$, at any location $r$ is given by (31):

$$q(r) = \frac{\omega}{2} \varepsilon_0 \, \text{Im}(\varepsilon) |E(r)|^2 \qquad (1)$$

where, $\varepsilon_0$ and $\varepsilon$ are the electric permittivity of the free space and the material respectively, $\omega$ is the frequency of illumination and $E(r)$ is the electric field at location $r$ inside the structure.

Since the gold plate is immersed in water, the heat generated is conducted to the nearby regions. We found the maximum temperature increase to be 7.81 K at the upper surface of the gold microplate. The temperature profiles in x-y plane and x-z plane are shown in figure 3(a), (c) respectively. Figure 3(b),(d) show the temperature distribution along the dotted lines in figure 3(a),(c) respectively. Supplementary figure S2 (a)-(c) show the temperature gradients set up in the region surrounding the gold microplate in the x, y and z directions. By running a time dependent study of the system, we calculated the evolution of the temperature of the water at a point 1 um above the surface of the gold microplate (Supplementary figure S3(d)). It reaches a steady state profile within few milliseconds. The maximum convective velocity of the fluid obtained is $3.40 \times 10^{-16}$ ms$^{-1}$ (see supplementary figure S4). We also calculated the velocity of few incoming particles through video analysis as they approach the gold microplate within first 30 s. The average velocity of the particles calculated was $1.18 \pm 0.44$ μms$^{-1}$. Thus we can conclude that the convection in the fluid is negligible and the dynamics is governed by thermophoresis rather than convection. This is consistent with the study reported previously (29).

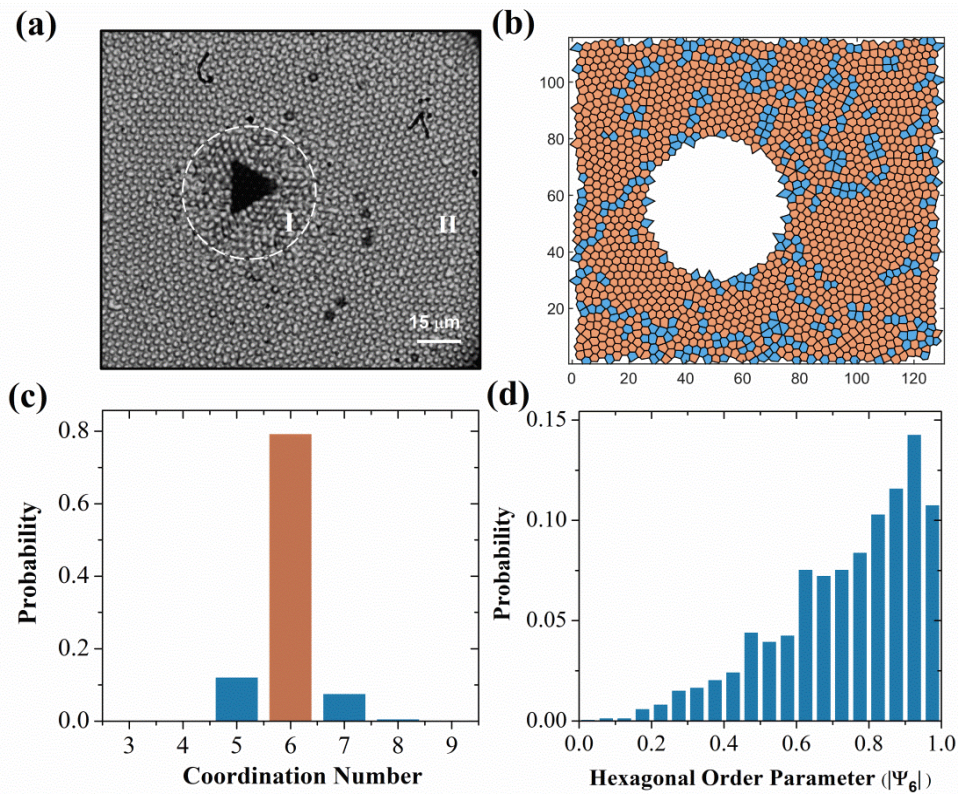

**Figure 4. (a)** Greyscale image of the assembly of dielectric particles (at 80 min), the region marked I marked by white dotted circle (diameter ~ 47.8 um) has been removed since the particles cannot be identified individually. Particles positions in the region II was tracked and fed into Voronoi analysis program. The Voronoi tessellation is shown in **(b)** where the hexagonal cells depicting Wigner-Seitz cells of particles having 6 nearest neighbours are colored in orange and the assembly regions with particles having nearest neighbour other than 6 is shown with blue cells. The normalized distribution of the coordination number of the particles is shown in **(c)**. **(d)** Hexagonal order parameter of the assembled particles. The distribution peaking around 0.95 depicts the closed hexagonal packing of the assembly.

### 3.2 Analysis of the assembly

As shown in figure 2 with increasing time, the scale of the assembly grows and reaches a saturation point, where the number of particles accumulated within a given area do not increase significantly over time. We take a snapshot of the assembly after an approximate saturation point (~ 80 min after laser is turned on) and characterize it. Characterization of the assembly before the formation of second layer around 45 minute is given in supplementary figure S5. Characterization of the assembly is imperative to the study since it helps us recover information such as local ordering. One of the ways to measure local ordering of the assembly is to calculate network of nearest neighbour of a particle for a given set of particle coordinates. Although there are many ways to calculate such networks (44), Voronoi decomposition (45, 46) is one of the unique method and its ease of application makes it particularly useful for our analysis. Voronoi decomposition pictorially maps out the extended 2D array of particles in the region of interest into number of polygons. Each polygon is a Wigner-Seitz cell for a particle, where the number of side represents the number of nearest neighbours adjacent to the particle. Counting the number of edges of each polygon also gives us a hint towards the symmetry of the assembled 2D array of particles. This is subjected to an accurate measurement of the particle coordinates, since a slight displacement from a preferred lattice position may lead to a large change in the coordination around the neighbourhood of a particular particle.

Upon close assessment of the assembly formation in figure 4(a), two distinct regions can be identified. Region I, close to the gold microplate (marked by a white dotted circle of diameter ~ 47.8 μm) where each particle cannot be individually identified due to multilayer formation; and region II outside the dotted circle, far away from the plate where apart from few defects, the assembly can be seen to be more uniform. We reject region I and analyze the assembly formed in region II. Particles in region II were mapped by trackmate (47). The region of interest contains approximately 1900 particles. The mapped particle coordinates were then fed into a modified Voronoi analysis program. The orange hexagonal cells in figure 4(b) represnt regions with particles having co-ordination number 6, whereas the blue cells show regions with co-ordination number other than 6. In our analysis, we have rejected the cells forming around the boundaries by using a modified Voronoi analysis (48) algorithm. It can be seen from the figure 4(b) that away from the boundaries, the hexagonal orange cells show more regular pattern compared to the peripheral cells. We plot the probability of a particle having coordination number from 3 to 8 in figure 4(c). It is evident that the majority of the assembly consists of particles with 6 nearest neighbours around them (marked with orange colored bin). Although particles may have six nearest neighbours around them, it doesn't necessarily assure hexagonal ordered packing. Hence, we further quantified hexagonal order parameter given by (4, 17):

$$\psi_6^j = \frac{1}{z_j}\sum_{m=1}^{z_j} \exp(i6\theta_m^j) \quad (2)$$

where, $z_j$ defines the number of nearest neighbour to the $j^{th}$ particle, m denotes the neighbours and $\theta_m^j$ is the angle between a reference axis and the line joining $j^{th}$ and $m^{th}$ particle.

As shown in figure 4(d), the histogram data peaks around 0.95, which indicates close hexagonal ordering of the assembled particles. The broad distribution of the histogram data can be attributed to the large area under consideration which includes the particles having nearest neighbours other than 6.

### 3.3 Effect of excitation polarization on the assembly

Since, the heat generated by a metallic structure essentially depends on the electric field distribution inside the structure,

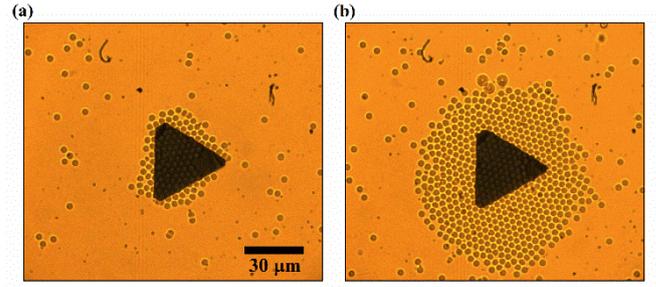

**Figure 5.** Polarization dependence of the microplate on the accumulation of the particles: **(a)** Accumulation of the particles at 25 min when the input excitation is s-polarized. **(b)** Assembly extends up to tens of microns at 25 min when the excitation is p-polarized.

we performed a polarization dependence study keeping the excitation power constant at 300 mW. We observed that the particles assembled for both s and p polarized light, but the extent of the assembly is different in both the cases. First, we kept the polarization of the excitation as s i.e. perpendicular to the plane of incidence. The snapshot is taken at 25 min as shown in figure 5(a). Then, we switched off the laser and allowed the particles to re-disperse into the solution. Now, we switched the polarization from s to p, i.e. parallel to the plane of incidence. When the polarization is kept at p, we observed that the particles experience a greater force of attraction towards the microplate. The particles assemble faster and also cover larger area in the same time period as can be seen by comparing figure 5(a) and figure 5(b). It can be explained on the basis of the difference in the amount of heat generated in both the cases. In order to generate heat in a structure, it is essential to excite the bulk of the volume (42). Compared to a p polarized light where the electric field can excite the bulk of the volume, s-polarized light is less efficient in generating heat. Thus, the polarization plays an important role in controlling the extent of the assembly and can be tuned according to the structure at hand.

### 3.4 Effect of surface to volume ratio of the microplate

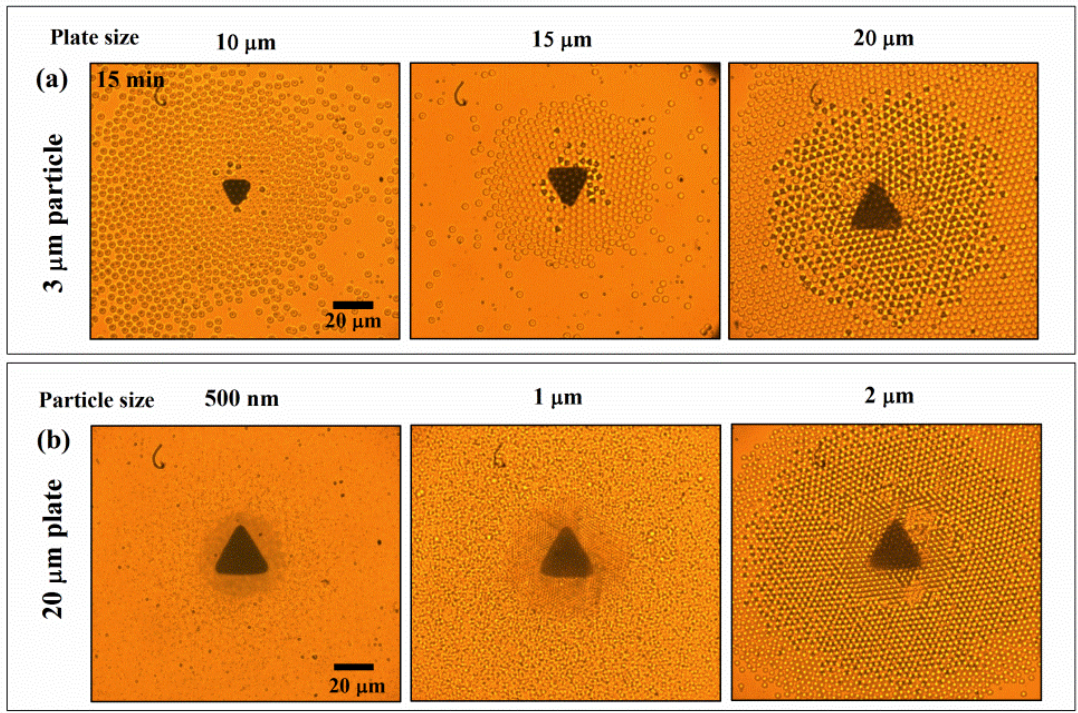

**Figure 6:** **(a)** for a fixed particle size of 3 *µ*m, the surface to volume ratio of the gold microplate is increased by keeping its volume fixed. As the surface area of the microplates increases from 10 *µ*m to 20 *µ*m, particles are more tightly bound to the plate and start accumulating at the second layer. **(b)** Keeping the size of the microplate constant, the size of the colloidal particles is varied from 500 nm to 3 *µ*m. The number of particles which are tightly bound to the gold microplate increases as the size of the particles is increased. All the snapshots are taken at 15 min.

Since, surface to volume ratio of a structure plays a crucial role in power dissipation and in turn efficient generation of heat, the size of the assembly can be tuned accordingly. Since efficient excitation of the volume of the structure is essential, to do a comparative study we kept the volume of all the structures same and varied the surface area. We lithographically prepared structures of sizes 10 μm (S/V: 14.28 μm$^{-1}$), 15 μm (S/V: 32.10 μm$^{-1}$) and 20 μm (57.14 μm$^{-1}$). The experiments were performed with 3 μm silica particles and the size of the plate was varied. The snapshot of the assembly at 15 min for every size of the gold microplate is shown in figure 6(a). The sizes of the structures used is tabulated in supplementary table 1. It can be seen from the snapshot that the efficiency of the assembly increases as we increase the surface area of the structure. For a 10 μm plate, the particles are loosely bound to the structure and display fluid like state. Very few particles near the structure can be seen forming the second layer. As we switch to 15 μm structure, the particles are more tightly bound to the structure and the second layer formation becomes prominent. Finally, when the S/V ratio is largest for 20 μm structure, the extent of the assembly is substantially larger. The particles are tightly bound to each other even at longer distances and the second layer formation extends up to 82 μm. Thus, the nature of assembly can be tuned to be either loosely or tightly bound, single or multilayer by tuning the surface to volume ratio of the structure.

### 3.5 Effect of particle size variation

Next, we kept the size of the structure constant and varied the particle size. The gold plate with largest surface to volume ratio (20 *µ*m) was chosen as it is the most efficient for assembly process. Colloidal particles of sizes 500 nm, 1 *µ*m, 2 *µ*m and 3 *µ*m were chosen for study. Upon laser excitation, there is a competitive behaviour between Brownian motion and thermophoresis. Figure 6(b) shows the snapshot of the assembly for different particle radii at 15 min. Particles of smallest size (500 nm) were tightly bound to the structure up to a distance of 34 *µ*m.

Rest of the particles were loosely bound to the gold microplate. For 1 $\mu m$ particles, the area which is tightly bound to the heat source increases and extends up 46 $\mu m$. As we further increased the size of the particles, diffusion decreased and more stable assembly was formed. 2 $\mu m$ particles showed multi-stacking of the particles. We can observe multiple layer formation up to 4 layers. Supplementary figure S6 indicates the number of layers formed. 4th layer formation can be seen in some regions in the vicinity of the gold microplate. 3rd and 2nd layers extend up to 45 $\mu m$ and 112 $\mu m$ respectively. Similar kind of assembly was also observed for 3 $\mu m$ particles as shown in last coloumn of Figure 6(a). This gives us an important insight regarding the optimization of the assembly by controlling the microstructure as well as particle size. Through the interplay of structure's surface to volume ratio and particle size, we can design the assembly process to achieve either single layer or multilayer assemblies.

### 3.6 Effect of composition on the assembly: Silica vs Polystyrene particles

Since, thermophoresis is sensitive to the compostion of the colloids (25), we investigated the accumulation with a particle of another material namely PS microspheres. A gold microplate was excited using 532 nm laser, and a solution containing 0.97 μm PS particles was injected on top of the microplate. We observed that few PS particles moved towards the gold microplate initially. There was no permanent accumulation of the particles near the gold microplate even after 30 min as shown in figure 7(a). Hence, we conclude that PS particles do not exhibit affinity towards the heat gradient as compared to silica particles. To compare the assembly process of PS and silica particles under the same heat gradient, solution containing 1 μm silica particles was injected into the sample. We observed that the silica particles assembled around the same microplate at the same input power as can be seen in figure 7(b). Although the PS particle has been shown to migrate away from the heat (33, 39), we did not observe such movement of the PS particles. This might be attributed to different particle-solvent interactions in our system. Thus, we conclude that thermophoretic force can play an important role in the assembly process.

The assembly of PS particles can be facilitated by excitation of an annealed gold thin film as reported in (49). We excited a gold film of thickness ~ 12 nm evanescently, and observed accumulation of both silica and PS particles. We attribute this assembly to the generation of thermoplasmonic field. A thin gold film aids the generation of surface plasmon polaritons (5, 50), facilitating the efficient assembly of PS particles. We can conclude that in our experiments, the PS particles do not exhibit affinity towards heat gradient and cannot be trapped by thermophoretic forces

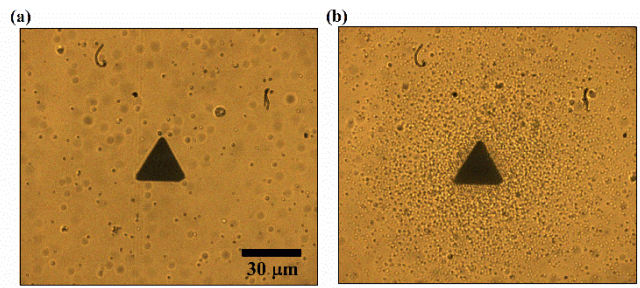

**Figure 7.** Comparison with polystyrene: **(a)** shows no accumulation of 0.97 μm polystyrene particles after 30 min . **(b)** 1 μm silica particles were added to the same solution. The accumulation of the particles starts after few seconds. The snapshot is taken after 30 min and shows the assembly of the silica particles.

alone, but an additional plasmonic field can assist in trapping PS particles.

### 4. Conclusion

We have experimentally studied the assembly of silica microparticles under the influence of an optothermal field of evanescently excited gold triangular microplate. Our study shows that the silica particles show reverse thermophoresis and can create colloidal assembly up to hundreds of microns. The packing order has been quantified in terms of co-ordination number and hexagonal order parameter which peaks around 0.95. We conclude that it is important to excite the volume of the structure for efficient generation of heat by varying surface to volume ratio as well as directing the polarization perpendicular to the structure. Whereas, silica particles show a favoured accumulation towards heat, we did not observe similar trend in the case of 0.97 μm polystyrene particles. This material selective assembly may have implications in colloidal sorting. Our system is advantageous over traditional optical tweezers as it is capable of creating large-scale assemblies and may complement the existing thermoplasmonic platforms. By using an optically-excited heat source, we can possibly utilize the thermal gradient around a metal microstructure which can be harnessed in opto-thermofluidic devices. A more elaborate study of the nature of the interacting forces that drive the colloids will lead to better understanding of light activated colloidal matter.

**Acknowledgements**

This work was partially funded by Air Force Research Laboratory and DST Energy Science (SR/NM/TP-13/2016) grant. Authors thank Adarsh B Vasista, Deepak K Sharma, Chetna Taneja for fruitful discussions and Rafeeque (Science and media center, IISER Pune) for drawing the schematic. GVPK thanks Dr. Guruswamy Kumaraswamy (NCL\IIT-B) and Dr. Apratim Chaterjee (IISER Pune) for fruitful discussion. GVPK acknowledges DST for Swarnajayanti fellowship grant (DST/SJF/PSA-02/2017-18).

# Supplementary Information

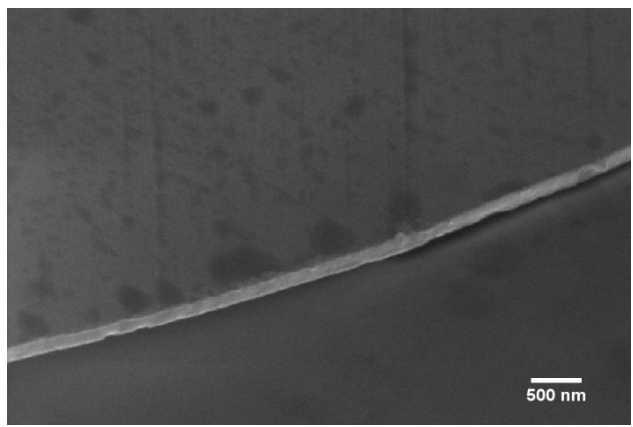

**Figure S3**: A representative image of the typical width of the edge of a microplate.

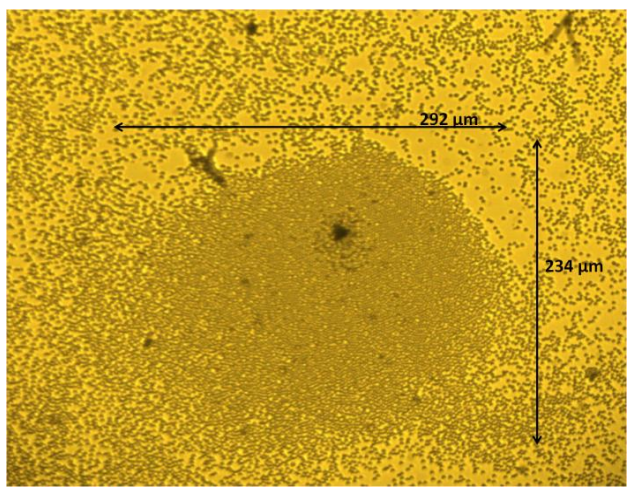

**Figure S2:** Long range assembly of 3 micron silica particles at 80 min.

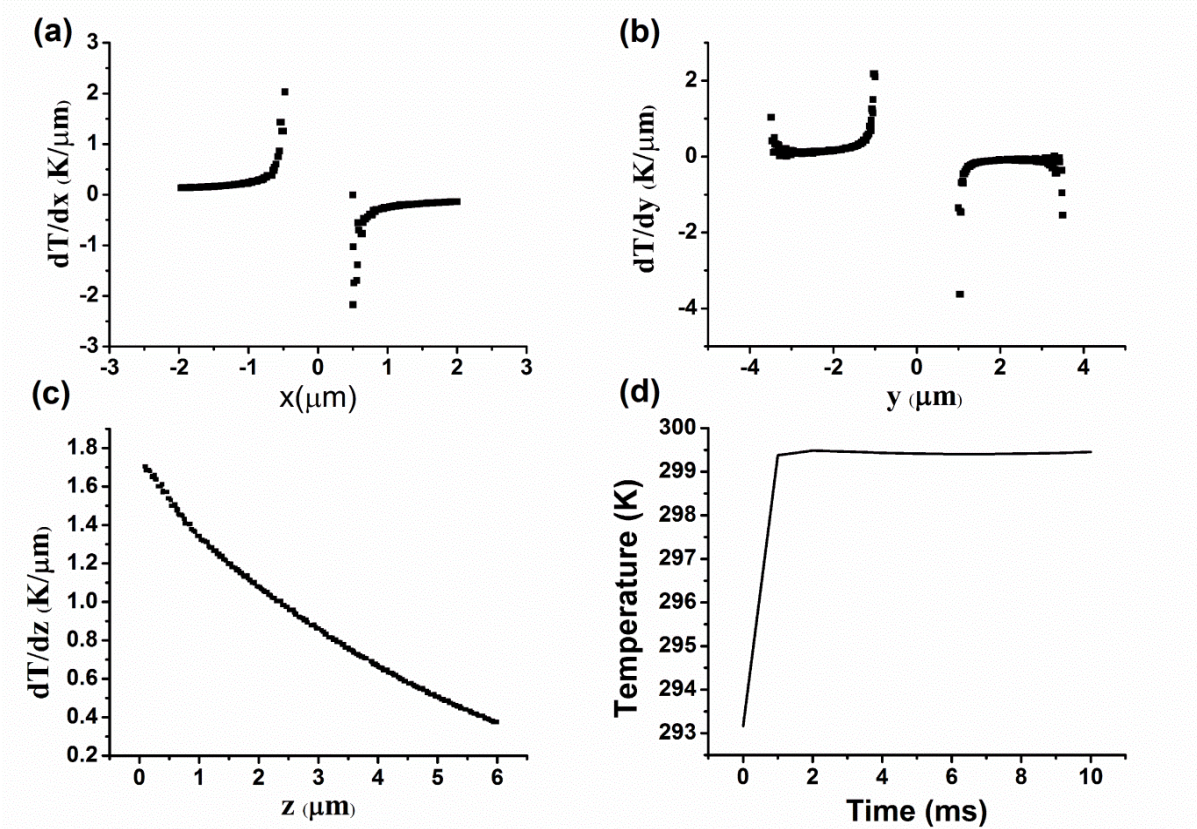

**Figure S3:** **(a)-(c)** Temperature gradients along x, y and z-directions are plotted. **(d)** Temperature of the water as a function of time at a distance 1 μm above the gold microplate.

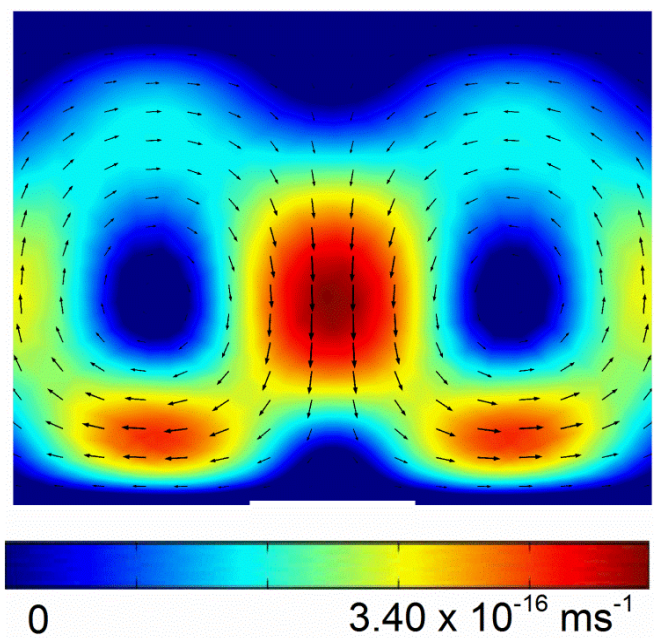

**Figure S4:** Convective velocity of the water above the surface of the gold microplate.

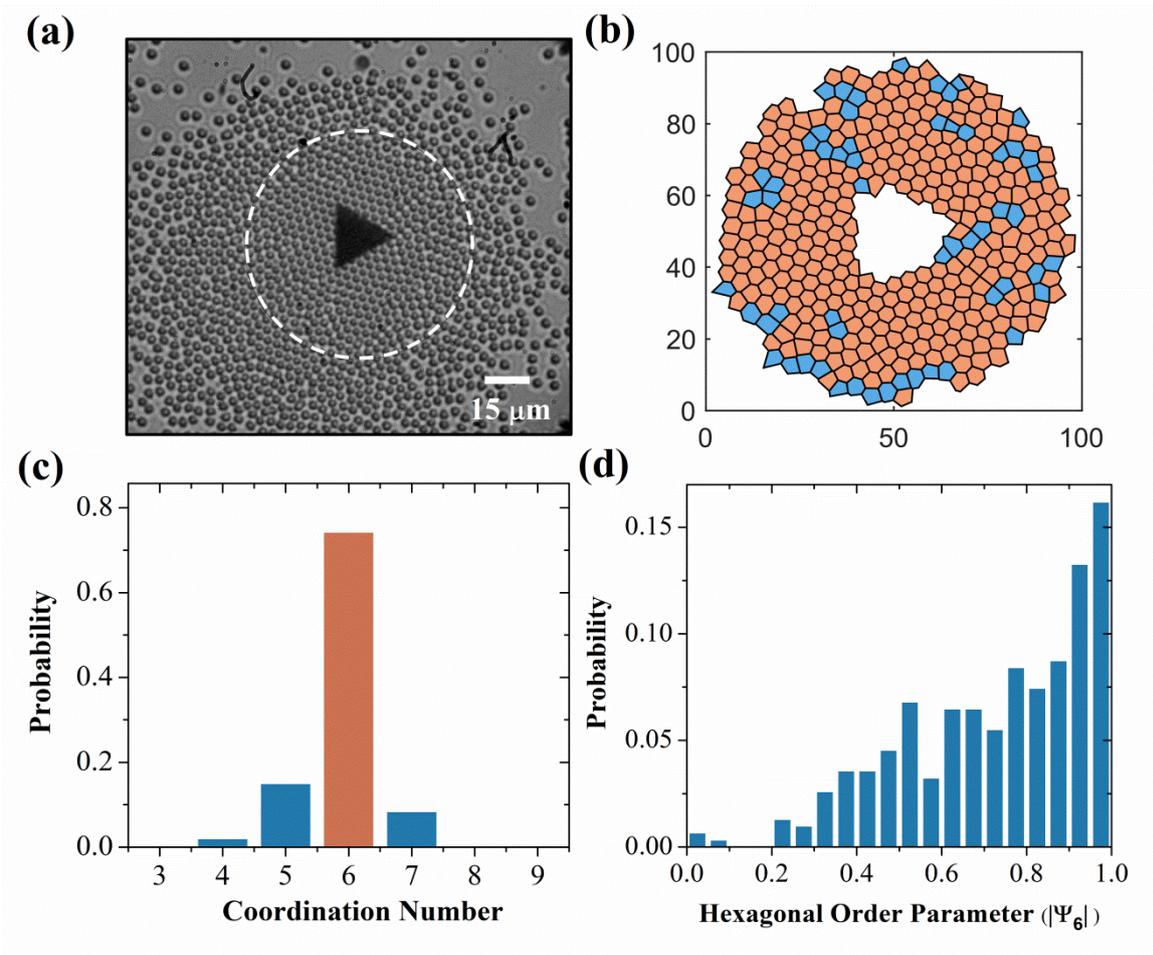

**Figure S5. (a)** Greyscale image of the assembly of dielectric particles (around 45 min). The region inside the dotted white circle was analysed. The positions of the particles were tracked and fed into voronoi analysis program. The Voronoi tessellation is shown in **(b)** where the hexagonal cells depicting Wigner-Seitz cells of particles having 6 nearest neighbours are colored in orange and the assembly regions with particles having nearest neighbour other than 6 is shown with blue cells. The normalized distribution of the coordination number of the particles is shown in **(c)**. **(d)** Hexagonal order parameter of the assembled particles. The distribution peaking around 1 depicts the closed hexagonal packing of the assembly.

| a ($\mu$m) | 10 | 15 | 20 |
|---|---|---|---|
| h (nm) | 70.0 | 31.1 | 17.5 |
| Area of the triangle S ($\mu m^2$) | 43.3 | 97.43 | 173.21 |
| Volume, V ($\mu m^3$) | 3.03 | 3.03 | 3.03 |
| S/V ($\mu m^{-1}$) | 14.28 | 32.1 | 57.14 |

**Table S1.** Different sizes of the gold microstructures prepared through lithography. Here, a represents each side of the microplate and h is its thickness.

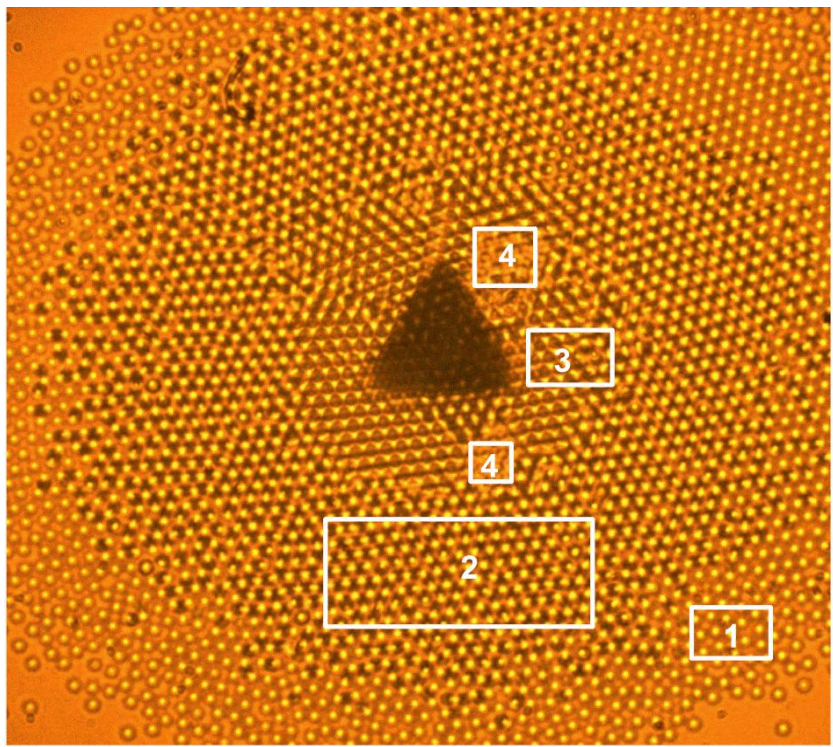

**Figure S6:** 2 um particles show multilayer formation for a 20 μm plate. The number in the box indicates the number of the layer that has formed.